\documentclass[aps,twocolumn,floatfix,superscriptaddress,prl]{revtex4-1}
\usepackage{amssymb}
\usepackage{graphicx}
\usepackage{dcolumn}
\usepackage{amsmath}
\usepackage{bm}
\usepackage{yhmath}
\usepackage{textcomp}
\usepackage{epstopdf}
\usepackage{color}
\usepackage{ulem}
\usepackage[toc,page,title,titletoc,header]{appendix}
\usepackage[colorlinks,
            linkcolor=blue,
            anchorcolor=blue,
            citecolor=blue
            ]{hyperref}
\usepackage{natbib}
\newcommand{\bfa}{{\bf a}}
\newcommand{\bfr}{{\bf r}}
\newcommand{\bfk}{{\bf k}}

\newcommand{\bfv}{{\bf v}}

\begin{document}

\title{Topological Condensate in an Interaction-induced Gauge Potential}
\author{Jun-hui Zheng}
\affiliation{Department of Physics, National Tsing Hua University, Hsinchu, Taiwan}
\author{Bo Xiong}
\affiliation{Department of Physics, National Tsing Hua University, Hsinchu, Taiwan}
\author{Gediminas Juzeli\={u}nas}
\affiliation{Institute of Theoretical Physics and Astronomy, Vilnius University, A. Go%
\v{s}tauto 12, Vilnius 01108, Lithuania}
\author{Daw-Wei Wang}
\affiliation{Department of Physics, National Tsing Hua University, Hsinchu, Taiwan}
\affiliation{Physics Division, National Center for Theoretical Sciences, Hsinchu, Taiwan}

\begin{abstract}
We systematically investigate the ground state and elementary excitations of a Bose-Einstein Condensate within a synthetic vector potential, which is induced by
the many-body effects and atom-light coupling. For a sufficiently strong inter-atom interaction, we find the condensate undergoes a Stoner-type ferromagnetic transition through the self-consistent coupling with the vector potential. For a weak interaction, the critical velocity of a supercurrent is found anisotropic due to the density fluctuations affecting the gauge field. We further analytically demonstrate the topological ground state with a coreless vortex ring in a 3D harmonic trap and a coreless vortex-antivortex pair in a 2D trap. The circulating persistent current is measurable in the time-of-flight experiment or in the dipolar oscillation through the violation of Kohn theorem.
\end{abstract}

\maketitle

\textit{Introduction}: Gauge fields play an important role in the modern particle physics, mediating interaction between elementary particles. In condensed matter physics, the gauge fields bring many important phenomena, such as integer/factional quantum Hall effects\cite{klitzing1980,tsui1982}, Laughlin liquids\cite{laughlin1983} and Hofstadter butterfly spectrum\cite{hofstadter1976}, etc. In quantum gas systems, artificial gauge fields can be also generated for neutral atoms in the rotating frame\cite{cooper2008,fetter2009}, or by the atom-light coupling with spatial dependent laser fields\cite{Juzeliunas2006,Juzeliunas2011,Juzeliunas2013} or detuning\cite{spielman2009,spielman20091}. The latter opens up a new possibility to study many-body physics with gauge potential, and is extended to investigate the spin-orbital coupling problems in similar experiments\cite{Lin2009prl}.

Besides of the generating by external lasers, it was theoretically proposed that the gauge field can be induced by dipole-dipole interaction between two dipolar or Rydberg atoms\cite{wenhui2013,wenhui20131,Cesa2013}. For many-body systems, some interesting dynamics\cite{juzeliunas20131} and excitations\cite{Edmonds2014} within a density- dependent gauge field were investigated in a 1D system. However, considering the finite temperature effects and quantum fluctuations, it is certainly more realistic and demanding to investigate many-body properties in higher dimensional systems, where the effective gauge field may further introduce topological defects and more interesting many-body physics not observable in 1D systems.

In this Letter, we systematically investigate the ground state and excitation properties of a (pseudo) spin-1/2 condensate with the interaction-induced gauge field in 2D and 3D systems. We find several new many-body properties: (i) For a sufficiently strong inter-atom interaction, the condensate can have a Stoner-type ferromagnetism through the self-consistent coupling with the synthetic field. (ii) In the weak interaction limit, we calculate the Bogoliubov excitation spectrum of a superfluid current, and show that the critical velocity for dynamical instability becomes anisotropic in space due to density fluctuations within the gauge field. (iii) In a harmonic trap, we show analytic solutions of the condensate ground state, and find a coreless vortex ring around the gauge field in a 3D trap and a coreless vortex-antivortex pair in a 2D trap. (iv) We further discuss how the topological condensate can be measured in the time-of-flight experiment and/or in the dipolar oscillation, where the Kohn theorem\cite{kohn1961, yip1991} fails due to the interaction-induced gauge field.

\textit{System Hamiltonian and its meanfield equation}: We consider bosonic atoms with two internal states subjected to a Raman coupling. Within the rotating wave approximation, the Hamiltonian can be expressed as $\hat{H}=\sum_{i=1}^N \hat{H}_1(i)+ \sum_{i<j}^N \hat{H}_2(i,j)$, where the single particle Hamiltonian is ($\hbar\equiv 1$)
\begin{equation}  \label{Hamiltonian}
\hat{H}_1= \frac{\mathbf{p}^2}{2m}\hat{\mathbf{I}}+\frac{1}{2}
\left[
\begin{array}{cc}
\Delta(\mathbf{r}) & \Omega\,e^{-2i\mathbf{k}_r\cdot\mathbf{r}} \\
\Omega\,e^{2i\mathbf{k}_r\cdot\mathbf{r}} & -\Delta(\mathbf{r})
\end{array}
\right]
+V_{\mathrm{trap}}(\mathbf{r})\hat{\mathbf{I}}
\end{equation}
with $\mathbf{k}_r$ being the recoil momentum and $V_{\mathrm{trap}}(\mathbf{r})$ being the spin-independent trapping potential. The two body interaction may be spin dependent and long-ranged in general, $H_2^{\sigma\sigma^{\prime}}= V_ {\sigma\sigma^{\prime }}(\mathbf{r}_i-\mathbf{r}_j)$,  with $\sigma,\sigma^{\prime } = \uparrow/\downarrow $ or $\pm$ being the pseudo-spin indices. Without losing of generality, we keep the spatial dependence in the detuning $\Delta(\mathbf{r})$ and take the Rabi frequency  $\Omega$ to be uniform in space. After  the second quantization, the total Hamiltonian becomes $\hat{H}= \sum_{\sigma,\sigma^{\prime }}\int d\mathbf{r} \hat{\Psi}_{\sigma}^\dag(\mathbf{r})H_1^{\sigma\sigma^{\prime }}\hat{\Psi}_{\sigma^{\prime }}(\mathbf{r})+\frac{1}{2}\sum_{\sigma,\sigma^{\prime }}\int d\mathbf{r} d\mathbf{r}^{\prime }\hat{\Psi}_{\sigma}^\dag(\mathbf{r})\hat{\Psi}_{\sigma^{\prime}}^\dag(\mathbf{r}^{\prime })V_{\sigma\sigma^{\prime }}(\mathbf{r}-\mathbf{r}^{\prime })\hat{\Psi}_{\sigma^{\prime }}(\mathbf{r}^{\prime })\hat{\Psi}_{\sigma}(\mathbf{r})$. Here $\hat{\Psi}_\sigma(\mathbf{r})$ is the bosonic field operator.

We will consider many-body properties at zero temperature in 2D and 3D systems,  for which the ground state $|G\rangle$ can be well-approximated by a coherent state wavefunction. As a result, the bosonic field operators satisfy: $ \hat{\Psi}_\sigma(\mathbf{r})|G\rangle = \Phi_\sigma(\bfr)|G\rangle$, where $\Phi_\sigma(\bfr)$ is the condensate wavefunction normalized to the total number of particles, i.e., $\sum_\sigma\int\Phi^\dag_\sigma(\mathbf{r})\Phi_\sigma(\mathbf{r}) d\mathbf{r}=N$. Applying the variational principle $\delta {\cal L}/ \delta{\Phi_\sigma}(\bfr) =0 $ to minimize the action $ {\cal L} =\langle G|\sum_\sigma\int\hat{\Psi}^\dag_\sigma(\mathbf{r})i \partial_t \hat{\Psi}_\sigma(\mathbf{r})d \bfr-\hat{H}|G\rangle $ , one arrives at the meanfield equation of motion (EOM)\cite{Jun2014} $~i \partial_t \Phi(\bfr)=\hat{H}_1 \Phi(\bfr) + \frac{1}{2}Q(\bfr)\hat{\mathbf{I}}  \Phi(\bfr) +  \frac{1}{2} G(\bfr) \hat{\sigma}_z \Phi(\bfr) $  for a wave-function of the spinor condensate, labeled by $\Phi(\bfr)\equiv [\begin{array}{c}\Phi_\uparrow(\mathbf{r}), \Phi_\downarrow(\bfr)\end{array} ]^T$ for conciseness. Note that the  equation of motion above is in form the same as a single particle equation with background fields: $Q(\mathbf{r})\equiv \sum_\sigma\int d\mathbf{r}^{\prime }\rho_\sigma(\mathbf{r}
^{\prime })[V_{\sigma\sigma}(\mathbf{r}-\mathbf{r}^{\prime })+V_{\sigma,-\sigma}(\mathbf{r}-\mathbf{r}^{\prime })]$  and $G(\mathbf{r})\equiv  \sum_\sigma \sigma\int d\mathbf{r}^{\prime }\rho_\sigma(\mathbf{r}^{\prime })[V_{\sigma\sigma}(\mathbf{r}-\mathbf{r}^{\prime })-V_{\sigma,-\sigma}(\mathbf{r}-\mathbf{r}^{\prime })]$.Here, $\rho_\sigma(\mathbf{r})\equiv |\Phi_\sigma(\mathbf{r})|^2$ is the number density of atoms in the spin $\sigma$ state. The effective detuning includes the many-body interaction through a meanfield shift: $\tilde{\Delta}(\mathbf{r})=\Delta (\mathbf{r})+G(\mathbf{r})$.

\textit{Interaction induced synthetic gauge field}:  In Ref.\cite{juzeliunas20131}, the  interaction-induced gauge  fields have been developed in the limit where the interaction is weak with respect to the Rabi frequency.  We begin our analysis by  presenting details of derivation beyond the perturbative regime. As we will show later,  the gauge field and dressed state are  mixed by a self-consistent equation, leading to  additional  interesting effects.  Following the standard process of adiabatic approximation, we firstly diagonalize the EOM without the kinetic energy by introducing a  spatially dependent transformation $S\equiv S(\mathbf{r})$:
\begin{equation}  \label{transferMatrix}
S=\frac{1}{\sqrt{2}}
\begin{bmatrix}
\mathrm{e}^{i \bfk_r\cdot\bfr}\sqrt{1+\tilde{\Delta}/\Lambda} & ~\mathrm{e}^{-i
\bfk_r\cdot\bfr}\sqrt{1-\tilde{\Delta}/\Lambda}~ \\
- \mathrm{e}^{i \bfk_r\cdot\bfr} \sqrt{1-\tilde{\Delta}/\Lambda} & ~ \mathrm{e}^{-i
\bfk_r\cdot\bfr}\sqrt{1+\tilde{\Delta}/\Lambda}~
\end{bmatrix},
\end{equation}
with $\Lambda \equiv\Lambda(\mathbf{r})=\sqrt{\tilde{\Delta}(\mathbf{r})^2+\Omega^2}$,  where  the $\mathbf{r}$-dependence will mostly be suppressed for writing convenience. Note that the transformation $S$ is density-dependent through the density dependence of the detuning $\tilde{\Delta}$.  As a result, the above EOM  is changed to $i \partial_t \tilde\Phi=\frac{(-i\nabla -i S \nabla S^\dag)^2}{2m} \tilde\Phi + [ V_{\mathrm{trap}} +\frac{Q}{2} + \frac{\Lambda}{2}\sigma_z ]\tilde\Phi$, where the transformed spinor wavefunction is $\tilde{\Phi}(\mathbf{r})\equiv [\tilde{\Phi}_+(\mathbf{r}),\tilde{\Phi}_-(\mathbf{r})]^T  \equiv S(\mathbf{r})\cdot {\Phi}(\mathbf{r})$.

In the transformed frame, the gauge field term is $-i S \nabla S^\dag = \frac{\Omega}{\Lambda}\mathbf{k}_r \sigma_x + \frac{\Omega \nabla \tilde{\Delta}}{2 \Lambda^2}\sigma_y-\frac{\tilde{\Delta}}{\Lambda}\mathbf{k}_r \sigma_z $, where the off-diagonal terms can be neglected within the adiabatic approximation.  Thus the above set of equations can be decoupled to the following effective equation for an individual component $ \tilde\Phi_\pm$\cite{Jun2014}:
\begin{eqnarray}  \label{adiabaticGPequation}
i\hbar\partial_t \tilde\Phi_\pm &=& \frac{\hbar^2}{2m}(-i\nabla \mp \frac{\tilde{
\Delta}}{\Lambda}\mathbf{k}_r)^2 \tilde\Phi_\pm +
V_{\mathrm{trap}}(\mathbf{r}) \tilde\Phi_\pm +\frac{Q}{2} \tilde\Phi_\pm \notag \\
&&  \pm \frac{\Lambda}{2} \tilde\Phi_\pm + \frac{\hbar^2}{8m}\frac{\Omega^2}{\Lambda^2}\left[
\frac{(\nabla \tilde{\Delta})^2}{\Lambda^2}+{4 \mathbf{k}_r^2}\right] \tilde\Phi_\pm.
\end{eqnarray}
We consider a generic case that the condensate is initially prepared in the lower branch ($\tilde{\Phi}_-$) state and remains there without involving the higher energy branch. Therefore, the total density is $\rho(\mathbf{r})=|\tilde\Phi_-(\mathbf{r})|^2$, and the density in original spin states are $\rho_\uparrow(\mathbf{r})=\left|S_{21}(\bfr)\right|^2 \rho(\mathbf{r})$ and $\rho_\downarrow(\mathbf{r})=\left|S_{22}(\bfr)\right|^2 \rho(\mathbf{r})$.

In the following, we concentrate on the short-ranged interaction case, i.e. $V_{\sigma\sigma^{\prime }}(\mathbf{r}-\mathbf{r}^{\prime })=g_{\sigma \sigma^{\prime}}\delta(\mathbf{r}-\mathbf{r}^{\prime })$ with $g_{\sigma \sigma^{\prime}}= 4 \pi \hbar^2 a_{\sigma \sigma^{\prime}}/m$, where $a_{\sigma \sigma^{\prime}}$ is the spin-dependent $s$-wave scattering length. After the unitary transformation into the new basis, the spin-dependent interaction energy $G(\mathbf{r})$ should be determined by the following self-consistent equation
\begin{equation}
G(\mathbf{r}) = \frac{\rho(\mathbf{r})}{2}\left[g_a-g_{as} \frac{\tilde{%
\Delta}(\mathbf{r})}{\Lambda(\mathbf{r})}\right], \label{shortG}
\end{equation}
where $\tilde\Delta(\bfr)\equiv\Delta(\bfr)+G(\bfr)$ and $\Lambda(\bfr)\equiv\sqrt{\tilde{\Delta}(\bfr)^2+\Omega^2}$ are also functions of $G(\bfr)$. Similarly, the spin-independent interaction energy is given by  $Q(\mathbf{r}) = \frac{\rho(\mathbf{r})}{2}[4 g_s-g_{a}\frac{\tilde{\Delta}(\mathbf{r})}{\Lambda(\mathbf{r})}]$. Here, $g_a \equiv g_{\upuparrows}-g_{\downdownarrows}$, $g_{as}\equiv g_{\upuparrows}+g_{\downdownarrows}-2g_{\uparrow\downarrow}$ and $g_{s}\equiv(g_{\upuparrows}+g_{\downdownarrows}+2 g_{\uparrow\downarrow})/4 $ \cite{Jun2014}.

\textit{General solutions}: A general solution for  the case of the short-ranged interaction can be obtained from the self-consistent equation  (\ref{shortG}).
Defining $\sinh\theta\equiv\tilde{\Delta}/\Omega$, Eq.(\ref{shortG}) becomes
\begin{eqnarray}
F[\theta]\equiv\sinh\theta+B\tanh\theta=C\,
\label{eq_theta}
\end{eqnarray}
with $B\equiv g_{as}\rho/2\Omega$ and $C\equiv(\Delta+g_a\rho/2)/\Omega$.

\begin{figure}[tbp]
\includegraphics[width=8.6 cm]{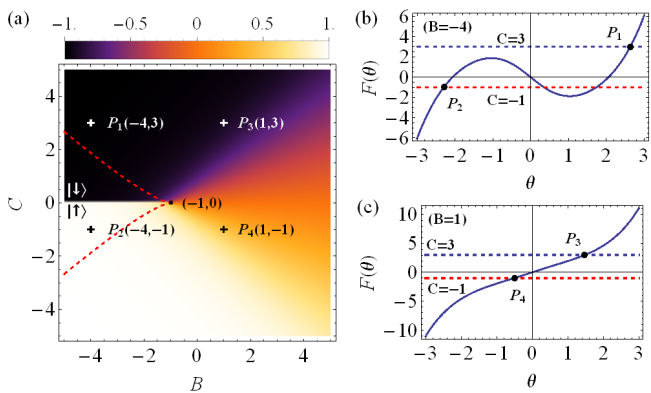}
\caption{(a) Density plot for ``magnetization'' ($M_\chi$) with $\Delta=0$ as a function of parameters $B$ and $C$ (see text). The regime between red dashed lines has three solutions of $\theta$ (Case II). (b) and (c) show the function $F[\theta]$ and the value of $C$ (horizontal dashed lines) for different points $P_{i}$ ($i=1,\cdots,4$) in (a). The values of $\theta$ for ground states of $P_{i}$ are denoted by filled circles.}
\label{phasediagram}
\end{figure}

Eq.(\ref{eq_theta}) always has at least one solution.  Approximate results can be obtained in different regimes (See Fig.\ref{phasediagram}): \textbf{Case I:} When $B>-1$, or when $B<-1 $ but $|C|>C_\ast$,  only one solution is available,  where $C_\ast\equiv F[\cosh^{-1}(|B|^{1/3})]$ is the critical value. In the regime of small $|C|$, the solution can be well approximated in the leading order expansion: $\theta\sim C/(1+B)=\frac{\Delta+g_a\rho/2}{\Omega+g_{as}\rho/2}$. As a result, the interaction energy ($Q$) includes higher order density fluctuation through the synthetic field. In the large $|C|$ regime, $\tanh\theta\sim \pm {\rm Sgn}(C)$, and we can obtain $\theta\sim \sinh^{-1}[C-B\text{Sgn}(C)]$. \textbf{Case II:} When $B<-1$ and $|C|< C_\ast$, there are three solutions and only one of them is the true ground state due to the interaction energy. We will discuss the ground state later. \textbf{Case III:} When $B<-1$ but $|C|=C_\ast$, there are two solutions at these two critical lines.

For  a uniform system with zero detuning ($\Delta=0$), Eq.(\ref{adiabaticGPequation}) shows that the ground state without persistent current has  a chemical potential: $\mu=(Q-\Lambda)/2+E_r\Omega^2/\Lambda^2= g_s\rho-\frac{1}{2}\Omega (C\tanh\theta+\cosh\theta)+ E_r/\cosh^2\theta$, where   $E_r \equiv \frac{k_r^2}{2m}$ is the recoiled energy. It is seen that the ground state prefers to select a solution of larger $|\theta|$. Consequently, in the regime of Case $\text{II}$,  the ``magnetization'' $ M_\chi \equiv (\rho_\uparrow-\rho_\downarrow)/\rho= -\tilde\Delta/\Lambda=-\tanh\theta$ is discontinuous at $C=0$ and $B<-1$. This indicates a first order Stoner-type quantum phase transition with a critical point (zero discontinuity) at $(B,C)=(-1,0)$. In Fig.\ref{phasediagram}, we show the value of $M_\chi$ in the $B-C$ diagram by setting $\Delta=0$. We stress that such  a phase transition merely occurs in a regime of rather strong interaction, i.e., $ g_{as} \rho \sim \Omega\gg E_r$, which is consistent with the adiabatic approximation\cite{Juzeliunas2011}.

\textit{Excitations and instability of superfluid current}:
Now we consider the condensate in a uniform space in the weak interaction ($|g_{a,as,s}| \rho(\mathbf{r})\ll \Omega$) and zero detuning ($\Delta(\mathbf{r})=0$) limit, which is corresponding to both $|B| \ll 1$ and $|C|\ll 1$ (i.e., \textbf{Case I}). From the analysis above, we have $G(\mathbf{r})=g_a\rho(\mathbf{r})/2 $ and $Q(\mathbf{r})=2g_s\rho(\mathbf{r})$ in the leading order, and the Gross-Pitaevskii(GP) equation is simplified to be (after neglecting the constant term and  setting $\psi(\mathbf{r})\equiv \tilde\Phi_-(\mathbf{r})$)
\begin{equation}  \label{gpequ}
i\hbar\partial_t \psi= \frac{\hbar^2}{2m} (-i\nabla+\mathbf{a}\rho)^2 \psi
+b\rho\psi,
\end{equation}
where $\mathbf{a}\equiv g_a \mathbf{k}_r/2 \Omega $ and $b\equiv g_s$.  This equation is consistent with the 1D result of Ref.\cite{juzeliunas20131},  yet the higher order terms would differ.

We first consider a general meanfield solution with a superfluid current: $\psi(\mathbf{r},t)=\sqrt{\rho_0}e^{im\mathbf{v}_g\cdot\mathbf{r}}e^{-i\mu
t}$, where $\mathbf{v}_g$ is the global phase gradient velocity. The resulting superfluid velocity, $\bfv_s\equiv \bfv_g+2\bfv_a$, also depends on the gauge field through the effective velocity, $2\bfv_a\equiv\bfa\rho/m$. We then apply a standard hydrodynamic approach to derive the elementary excitation dispersion\cite{Jun2014}:
\begin{eqnarray} \label{velocity}
\omega_\bfk &=&\mathbf{k}\cdot\left(\mathbf{v}_{s}+\bfv_a\right)+
\nonumber\\
&&+\sqrt{(c^2+2\bfv_s\cdot\bfv_a
+(\bfv_a\cdot\hat{k})^2)\bfk^2+\frac{\mathbf{k}^{4}}{4m^{2}}},
\label{w_k}
\end{eqnarray}
where $c\equiv\sqrt{b\rho_0/m}$ is  a standard Bogoliubov phonon velocity without  including the gauge field.

In the regime of long wavelength and absence of the gauge field ($\bfv_a=0$), we have $\omega_\bfk=\bfv_s\cdot\bfk+c|\bfk|+{\cal O}(|\bfk|^2)$, which shows the standard Landau critical velocity of a stable supercurrent: $|\bfv_s|<c$. When the gauge field is finite, we introduce a specific coordinate with the gauge field along the $x$ axis, i.e. $\bfv_a=v_a (1, 0, 0)$, and then $\bfv_s=v_s (\cos\xi,\sin \xi,0)$, and $\bfk= k(\sin\gamma \cos \kappa, \sin\gamma \sin \kappa, \cos \gamma)$. Thus, we obtain $v_{\hat{k}} \equiv \lim_{|\bfk|\rightarrow 0} {\omega_{\bfk}}/{|\bfk|}=v_{s}\sin \gamma \cos \left( \kappa -\xi \right)
+v_{a}\sin \gamma \cos \kappa +\sqrt{c^{2}+2v_{s}v_{a}\cos \xi +v_{a}^{2}\sin^{2}\gamma \cos ^{2}\kappa}$. The condition of a stable superfluid requires
$v_{\hat{k}}>0$ for $\gamma\in \left[ 0,\pi \right]$, and $\kappa,\xi\in \left[ 0,2\pi \right)$.

\begin{figure}[tbp]
\includegraphics[width=7 cm]{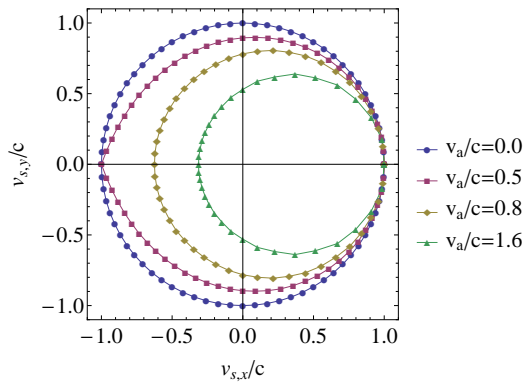}
\caption{The critical velocity for different ratio $v_a/c$, where $\bfv_a$ is along the x-axis. The superfluid is stable inside the circle.}
\label{vel}
\end{figure}

In Fig.\ref{vel}, we show the calculated critical superfluid velocity, $(v_{s,x}/c, v_{s,y}/c)$ (results in $y$ and $z$ directions are the same). As  one can see, the critical velocity becomes different for the superfluid current flowing in different direction with respect to the gauge field. In fact, in the weak interaction limit with $v_a\ll c$, the minimum critical velocity appears neither along nor perpendicular to $\bfv_a$, showing a hybridization effect of the synthetic gauge field and superfluidity. This results from the density fluctuation of the gauge field in the kinetic energy of Eq.(\ref{gpequ}),  so the effect does not exist if the synthetic gauge field is generated by the laser field only.

\textit{Ground state in a harmonic trap}:
Now we solve the static GP equation of (\ref{gpequ}) with a harmonic trap $V_{\rm{trap}}(\bfr)= \frac{1}{2} m \omega^2 {\mathbf{r}}^2$. For a 1D trap along the $x$-axis considered in Ref. \cite{juzeliunas20131}, the gauge field can be eliminated via the transformation: $\psi(x)=\psi_0(x) \exp\left[-i \int_0^x a \rho (x^{\prime }) dx^{\prime }\right]$. The transformed wavefunction $\psi_0(x)$ satisfies the  conventional GP equation without the gauge field term and therefore can be well-approximated within the Thomas-Fermi approximation (TFA). As a result, the  ground state wavefunction is
$\psi(x)= \sqrt{\frac{\mu-m \omega^2 x^2/2}{b}} \exp [-i\frac{a(\mu
x -m \omega^2 x^3/6)}{b}]$.
Note that the particle current is still zero after including the gauge field, even though the wavefunction is complex.

For higher dimensional system, the gauge field cannot be completely gauged away.  Setting $\psi=\sqrt{\rho}e^{i\eta}$, we start from the following static hydrodynamic equations:
\begin{subequations}
\label{allequations} 
\begin{eqnarray}
0 &=& -\nabla \cdot \mathbf{j}_s =-\nabla \cdot (\rho \mathbf{v}_s),
\label{phase} \\
\mu &=& \frac{-\hbar^2}{2 m \sqrt{\rho}}\nabla^2 \sqrt{\rho} +\frac{1}{2}m
\mathbf{v}_s^2 +b \rho +\frac{1}{2} m \omega^2 \mathbf{r}^2,  \label{density}
\end{eqnarray}
\end{subequations}
where $\mathbf{j}_s$ is the current density and $\mathbf{v}_s(\mathbf{r})=\frac{1}{m} (\nabla \eta + a \rho\hat{x})$ is the superfluid velocity. We
can get density function from Eq.(\ref{density}) within TFA again by neglecting the kinetic energy term, so that $\mu=m \omega^2 R^2/2$ and $\rho=m \omega^2 (R^2-r^2)/2b$. Here $R$ is the TF radius and determined by the total number of particles. Substituting this density expression in Eq.(\ref{phase}), the differential equation for $\eta$ can be obtained \cite{Jun2014}:
\begin{equation}  \label{phaseeqn}
-\frac{ (R^2-\mathbf{r}^2)}{2} \nabla^2 \eta + r
\partial_r \eta = f(\mathbf{r}),
\end{equation}
where $f(\mathbf{r}) \equiv -{a m \omega^2 } x (R^2-\mathbf{r}^2)/{b}$. In 2D system, its homogeneous equation, $-\frac{(R^2-\mathbf{r}^2)}{2} \nabla^2 \eta + r \partial_r \eta =\epsilon \eta$, has eigenstate solutions in  the polar coordinate $(r,\theta)$ to be $\eta_{nl}=\tilde{C}_{nl} r^{|l|} {}_2 F_1(-n,|l|+n+1,|l|+1; \frac{r^2}{R^2}) e^{-i l \theta}$\, with eigenvalues $\epsilon_{nl}=|l|+2n+2n|l|+2n^2$. Here $\tilde{C}_{nl}$ is the normalization coefficient, ${}_2 F_1$ is  the hypergeometric function and $n$ is a non-negative integer.

Equation (\ref{phaseeqn}) can then be solved by Green's function method and the obtained the solution is
\begin{equation}  \label{phase3dsolution}
\eta(\mathbf{r})= \sum_{n,l} \int d\mathbf{r}^{\prime }\frac{\eta_{nl}(\mathbf{r%
})\eta_{nl}^*(\mathbf{r^{\prime }})}{\epsilon_{nl}} f(\mathbf{r}%
^{\prime }) = \frac{a  m \omega^2 x (r^2-3 R^2) }{7b} , \notag
\end{equation}
because only a few components give non-zero contribution within such the TF approximation. As a result, the superfluid velocity distribution is
\begin{eqnarray}  \label{velocity2d}
\mathbf{v}_{s} &=& \frac{2 a \omega^2}{7 b} \left(x^2 + R^2/4-5r^2/4, ~~x y\right).
\end{eqnarray}

\begin{figure}[tbp]
\includegraphics[width=8.5 cm]{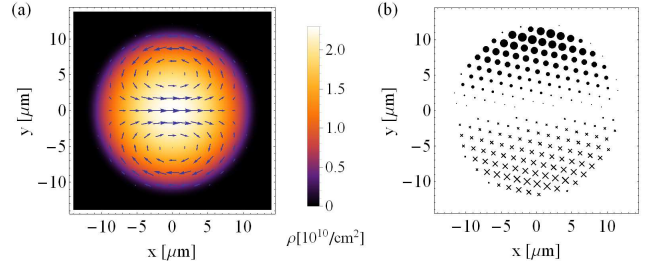}
\caption{ (a) The particle density (brightness) and current-density (arrows)
distribution in a 2D harmonic trap. (b)The density dependent magnetic field $\mathbf{B}=\nabla \times \bf{a} \rho$ induced by interaction. Here we use parameters of ${}^{87}\mathrm{Rb}$ atoms with the two spin states, $|\uparrow \rangle=|F=2, m_F=+1 \rangle$ and  $|\downarrow \rangle=|F=1, m_F=-1\rangle$ \cite{Dan2013}, trapped in a quasi-2D harmonic trap with trapping frequency $(\omega, \omega)=2 \pi \times (60, 60)~ \mathrm{Hz}$. Total particle number is $N=5.0 \times 10^4$, the recoiled momentum is $k_r =2\pi\times 1 \mu\mathrm{m}^{-1}$, and the Raman coupling is $\Omega= 2 \pi \times 10 \mathrm{kHz}$.  Large scattering lengths, $a_a= 3\mathrm{nm}$ and $a_s=6\mathrm{nm}$, are used when near the Feshbach resonance\cite{Fedichev1996,Theis2004,Enomoto2008,Papoular2010}. The chemical potential is $\mu \simeq 2\pi \times 2.1  \mathrm{kHz}$ for a transverse confinement length 0.53 $\mu$m.
}
\label{corelessvortex}
\end{figure}

In Fig.\ref{corelessvortex}(a), we show the full numerical results of the particle density and the current density distribution, which agree with the
analytic results (not shown) very well. We note that the ground state has a pair of coreless vortexes for the persistent current, which is the most important
feature of the interaction  induced synthetic gauge field. The center of vortexes (given by the zero velocity point)  is located at $(0, \pm R/ \sqrt{5})$. We emphasize that such  a coreless vortex can be regarded as a result of a synthetic magnetic field flux, which is induced by the synthetic gauge field (Fig.\ref{corelessvortex}(b)).

For a condensate in a 3D symmetric trap, we can follow similar approach and obtain the phase $\eta(\mathbf{r})= a  m \omega^2 x (r^2-3 R^2) /8b$, so that the superfluid velocity is
\begin{eqnarray}  \label{velocity3d}
v_{sx} &=& \frac{a \omega^2}{4 b}  x^2+\frac{a \omega^2}{8 b} (R^2-3r^2), \\
v_{sy} &=& \frac{a \omega^2}{4 b} x y,~~~v_{sz} = \frac{a \omega^2}{4 b} x z,
\end{eqnarray}
where the persistent current forms a coreless vortex ring with  a radius $R/\sqrt{3}$ \cite{Jun2014}.

\textit{Magnetic properties in spin basis}: Using the inverse transformation of (\ref{transferMatrix}), we can express the ground state back to the
original spin basis, $\Phi_\uparrow(\mathbf{r})=S_{21}(\mathbf{r})^*\phi(\mathbf{r})=-\sqrt{\rho(\mathbf{r})}\sin\beta(\mathbf{r})\,e^{i\eta(\mathbf{r})-i\mathbf{k}_r\cdot\mathbf{r}}$ and $\Phi_\downarrow(\mathbf{r})=S_{22}(\mathbf{r})^*\phi(\mathbf{r})=\sqrt{\rho(\mathbf{r})}\cos\beta(\mathbf{r})\,e^{i\eta(\mathbf{r})+i\mathbf{k}_r\cdot\mathbf{r}}$, where $\cos\beta=\sqrt{1+\tilde{\Delta}(\mathbf{r})/\Lambda(\mathbf{r})}/\sqrt{2}$. It is
shown that the spin density becomes $S_z(\mathbf{r})\equiv \Phi^\dag(\bfr) \frac{\sigma_z}{2} \Phi(\bfr)/\rho(\bfr)=-\frac{1}{2}\tilde{\Delta}(\mathbf{r})/\Lambda(\mathbf{r}) $ and $S_+(\mathbf{r})\equiv \Phi^\dag(\bfr) \frac{\sigma_x+i\sigma_y}{2} \Phi(\bfr)/\rho(\bfr) =-[\Omega/2\Lambda(\mathbf{r})]\,e^{i 2\mathbf{k}_r\cdot\mathbf{r}}  $ which in general gives a canted spiral phase with a period of $\pi/k_r$. However, since the phase $\eta(\mathbf{r})$ is the same for both spin  components, the density measurement in real space cannot reveal the vortex structure displayed here.

\begin{figure}[tbp]
\includegraphics[width=8.5 cm]{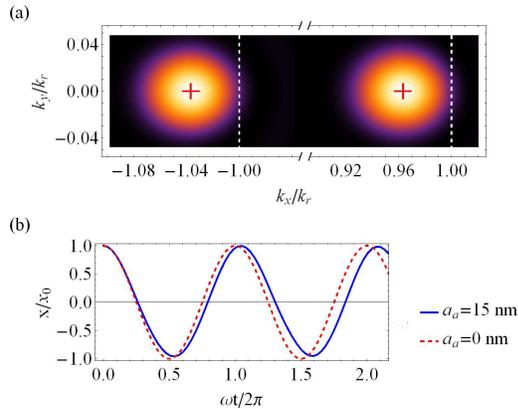}
\caption{(a) Momentum distribution of a typical condensate shown in Fig.\ref{corelessvortex}, which can be measured in the TOF experiment. Left/right part is the spin up/down component. The vertical dashed line is guiding the position of condensate center without gauge field. (b) The dipolar oscillation of a condensate of total $ 5 \times 10^3$ particles with an initially displaced center at position $(x_0=2.8 \mu m, 0)$ in a 2D trap of frequency $2\pi\times(240, 240)$ Hz. To signify this effect induced by gauge field, we use the parameter $a_a=15 \mathrm{nm}$ (solid
/blue line), 
while other parameters are the same as used in Fig. \ref{corelessvortex}.
}
\label{TOF}
\end{figure}

\textit{Experiment measurement}: There are several methods to measure the topological condensate proposed in this paper. We first consider time-of-flight (TOF) measurement, where the two spin components are decoupled and freely expand. In Fig.\ref{TOF}(a), we show the momentum distribution calculated from the condensate in a 2D trapped system.  The condensate cloud of the two species  is clearly seen to have a constant shift in the momentum distribution with respect to the their recoiled momentum, $\pm k_r$. The shift for each component can be easily calculated to be $\Delta \bfk_\uparrow\approx\Delta \bfk_\downarrow\approx \frac{\Delta N} { N} \bfk_r $ (if $\mu\ll\Omega$), where $\frac{\Delta N }{N}\equiv \frac{N_\uparrow-N_\downarrow}{N} = -\frac{g_a}{3g_s}\frac{\mu}{\Omega}= -\frac{g_a \rho(0)}{3 \Omega}  $ for the 2D case, and $ \frac{\Delta N }{ N}=-\frac{2g_a}{7g_s}\frac{\mu}{\Omega}=-\frac{2g_a \rho(0)}{7\Omega}$ for the 3D case within the TFA.  Here $\rho(0)$ is the density in the center of the trap. Note that the interaction induced gauge field can be enhanced by considering Feshbach resonance or confinement resonance.  The coreless vortices or vortex ring can be also measurable by interfering with another condensate of no (or different) gauge field.

In the presence of a uniform magnetic field, the celebrated Kohn theorem \cite{kohn1961, yip1991}  shows that the center-of-mass oscillation keeps the same frequency as the noninteracting one, independent of the inter-particle interaction. However, in the density dependent synthetic gauge field we discuss here, the effective magnetic field must be non-uniform, so that the Kohn theorem fails to apply, and dipolar oscillation can then have a frequency different from the noninteracting results. In Fig.\ref{TOF}(b), we show how the oscillation frequency along $\hat{x}$-direction is changed in the presence of the synthetic gauge field.

\textit{Summary}: In this letter, we have systematically derived a
generalized synthetic gauge field theory for a coupled two component bosonic system, where the inter-particle interaction is included by a self-consistent equation. In the strong interaction regime, we show a Stoner-type ferromagnetism by the interplay of gauge field and inter-particle  coupling. In the weak interaction limit, we show  the anisotropic critical velocity in a uniform space and the topological structure of  the persistent currents in a harmonic trap. These  distinctive features can be observed in the standard TOF and dipolar oscillation experiments.

We thank fruitful discussion with M. Cazalilla, Y.-J. Lin, Y.-C. Cheng, and C. S. Chu. This work is supported by MoST and NCTS in Taiwan,  as well as the project TAP LLT 01/2012 of the Research Council of Lithuania.

\bibliographystyle{apsrev4-1}

\end{document}


\title{Supplementary Material for\\
``Topological Condensate in an Interaction-induced Gauge Potential''}
\author{Jun-hui Zheng}
\affiliation{ Department of Physics, National Tsing Hua University, Hsinchu, Taiwan}
\author{Bo Xiong}
\affiliation{ Department of Physics, National Tsing Hua University, Hsinchu, Taiwan}
\author{Gediminas Juzeli\={u}nas}
\affiliation{Institute of Theoretical Physics and Astronomy, Vilnius University, A. Go%
\v{s}tauto 12, Vilnius 01108, Lithuania}
\author{Daw-Wei Wang}
\affiliation{ Department of Physics, National Tsing Hua University, Hsinchu, Taiwan}
\affiliation{Physics Division, National Center for Theoretical Sciences, Hsinchu, Taiwan}

\maketitle

\section{Gross-Pitaevskii Equation with Adiabatic Approximation}

\label{appendix1}

In this section, we devolep the Gross-Pitaevskii (GP) equation with
adiabatic approximation. Using the Ansatz for the ground state of the system being a coherent state $|G\rangle$, the field operators satisfy $ \hat{\Psi}_\sigma(\mathbf{r})|G\rangle = \Phi_\sigma(\bfr) |G\rangle $. Then the mean field energy $E \equiv \langle G|\hat{H}|G \rangle$ reads
\begin{eqnarray}
E &=&\sum_{\sigma \sigma ^{\prime }}\int d \mathbf{r}\Phi _{\sigma }^{\ast
}H_{1}^{\sigma \sigma ^{\prime }}\Phi _{\sigma ^{\prime }}  \notag \\
&&+\frac{1}{2}\int d \mathbf{r} d \mathbf{r}^{\prime }\left\vert \Phi
_{\uparrow }\left( \mathbf{r}\right) \right\vert ^{2}\left\vert \Phi
_{\uparrow }\left( \mathbf{r}^{\prime }\right) \right\vert ^{2}V_{\uparrow
\uparrow }(\mathbf{r}_\Delta)  \notag \\
&&+\frac{1}{2}\int d \mathbf{r} d \mathbf{r}^{\prime }\left\vert \Phi
_{\downarrow }\left( \mathbf{r}\right) \right\vert ^{2}\left\vert \Phi
_{\downarrow }\left( \mathbf{r}^{\prime }\right) \right\vert
^{2}V_{\downarrow \downarrow }(\mathbf{r}_\Delta)  \notag \\
&&+\int d \mathbf{r} d \mathbf{r}^{\prime }\left\vert \Phi _{\uparrow
}\left( \mathbf{r}\right) \right\vert ^{2}\left\vert \Phi _{\downarrow
}\left( \mathbf{r}^{\prime }\right) \right\vert ^{2}V_{\uparrow \downarrow }(%
\mathbf{r}_\Delta),  \label{energy}
\end{eqnarray}
where $\mathbf{r}_\Delta=\mathbf{r}-\mathbf{r}^{\prime}$ and
\begin{equation}  \label{Hamiltonian}
\hat{H}_1= \frac{\mathbf{p}^2}{2m}\hat{\mathbf{I}}+\frac{1}{2} \left[%
\begin{array}{cc}
\Delta(\mathbf{r}) & \Omega\,e^{-2i\mathbf{k}_r\cdot\mathbf{r}} \\
\Omega\,e^{2i\mathbf{k}_r\cdot\mathbf{r}} & -\Delta(\mathbf{r})%
\end{array}
\right]+V_{\mathrm{trap}}(\mathbf{r})\hat{\mathbf{I}}.
\end{equation}

By requiring the variation of the action  with respect to $\Phi_\sigma(\bfr)$  to be zero, i.e., $ \frac{\delta}{\delta \Phi_\sigma} \langle G|\sum_\sigma i \hat{\Psi}_\sigma^* \partial_t  \hat{\Psi}_\sigma-\hat{H}|G \rangle=0$,  we arrive  at a set of two GP equations,
\begin{equation}  \label{Ad2}
i\partial _{t}\binom{\Phi _{\uparrow }}{\Phi _{\downarrow }}=H_{1}\binom{%
\Phi _{\uparrow }}{\Phi _{\downarrow }}+\left(
\begin{array}{cc}
G_{1} & 0 \\
0 & G_{2}%
\end{array}%
\right) \binom{\Phi _{\uparrow }}{\Phi _{\downarrow }},
\end{equation}%
where
\begin{eqnarray}
G_{1}\left( \mathbf{r}\right) &=& \int d \mathbf{r}^{\prime }\left[%
\left\vert \Phi _{\uparrow }\left( \mathbf{r}^{\prime }\right) \right\vert
^{2}V_{\uparrow \uparrow }(\mathbf{r}_\Delta)+\left\vert \Phi _{\downarrow
}\left( \mathbf{r}^{\prime }\right) \right\vert ^{2}V_{\uparrow \downarrow }(%
\mathbf{r}_\Delta)\right],  \notag \\
G_{2}\left( \mathbf{r}\right)&=& \int d \mathbf{r}^{\prime }\left[\left\vert
\Phi _{\downarrow }\left( \mathbf{r}^{\prime }\right) \right\vert
^{2}V_{\downarrow \downarrow }(\mathbf{r}_\Delta)+\left\vert \Phi _{\uparrow
}\left( \mathbf{r}^{\prime }\right) \right\vert ^{2}V_{\uparrow \downarrow }(%
\mathbf{r}_\Delta)\right].  \notag
\end{eqnarray}
 Expressing the matrix $\text{diag}(G_{1,}G_{2})$   a linear combination of Pauli matrices, Eq.(\ref%
{Ad2}) becomes
\begin{equation}
i \partial _{t}\binom{\Phi _{\uparrow }}{\Phi _{\downarrow }}=H_{1}\binom{%
\Phi _{\uparrow }}{\Phi _{\downarrow }}+\left( \frac{Q}{2}+\frac{G}{2}\sigma
_{z}\right) \binom{\Phi _{\uparrow }}{\Phi _{\downarrow }},
\end{equation}%
with
\begin{equation}  \label{Ad4}
Q=G_{1}+G_{2},~~G=G_{1}-G_{2}.
\end{equation}%
 Thus the single-particle-like effective Hamiltonian is
\begin{eqnarray} \label{gpeeff}
H_{\text{eff}}\equiv H_{1}&+&\frac{Q}{2}+\frac{G}{2}\sigma _{z} \notag \\
=\Big[\frac{\mathbf{p}^{2}}{2m} &+& V_{\text{trap}}+\frac{Q}{2}\Big] \hat{%
\mathbf{I}}+\frac{1}{2}
\begin{bmatrix}
\tilde{\Delta}(\mathbf{r}) & \Omega\,e^{-2i\mathbf{k}_r\cdot\mathbf{r}} \\
\Omega\,e^{2i\mathbf{k}_r\cdot\mathbf{r}} & -\tilde{\Delta}(\mathbf{r})%
\end{bmatrix}
\end{eqnarray}%
with an effective detuning shifted by the interaction,
\begin{equation}
\tilde{\Delta}(\mathbf{r})=\Delta(\mathbf{r}) +G(\mathbf{r}).
\end{equation}

Performing the transformation of the wavefunction
\begin{equation}  \label{Ad7}
\tilde{\Phi} \equiv\left[\tilde\Phi _{+},\tilde\Phi _{-}\right] ^{T} = S\Phi
\end{equation}%
with the unitary matrix
\begin{equation}\label{Ad77}
S=\frac{1}{\sqrt{2}}
\begin{bmatrix}
e^{i\mathbf{k}_r \cdot \mathbf{r}}\sqrt{1+\tilde{\Delta}/\Lambda } & ~~ e^{-i
\mathbf{k}_r\cdot \mathbf{r}}\sqrt{1-\tilde{\Delta}/\Lambda } ~\\
-e^{i\mathbf{k}_r\cdot\mathbf{r}}\sqrt{1-\tilde{\Delta}/\Lambda } & ~~ e^{-i
\mathbf{k}_r\cdot \mathbf{r}}\sqrt{1+\tilde{\Delta}/\Lambda }
\end{bmatrix}
\end{equation}
 and $\Lambda =\sqrt{\tilde{\Delta}^{2}+\Omega ^{2}}$, one can diagonalize the second  term in the last line of Eq.(\ref{gpeeff}). In this new basis, the GP equation becomes
\begin{eqnarray} \label{equnew}
i\left( \partial _{t}+S\partial _{t}S^{\dag }\right) \tilde{\Phi} &=&\frac{1%
}{2m}\left( -i\nabla -iS\nabla S^{\dag }\right) ^{2}\tilde{\Phi} \notag\\
&&+\left(V_{\text{trap}}+\frac{Q}{2}+ \frac{\Lambda }{2}\sigma _{z}\right)
\tilde{\Phi}.
\end{eqnarray}%
Direct calculation shows,
\begin{equation} \label{vector}
-iS\nabla S^{\dag }=\frac{\Omega}{\Lambda }\mathbf{k}_{r} \sigma _{x}+\frac{%
\Omega \nabla \tilde{\Delta}-\tilde{\Delta}\nabla \Omega }{2\Lambda ^{2}}%
\sigma _{y}-\frac{\tilde{\Delta}}{\Lambda }\mathbf{k}_{r} \sigma _{z}.
\end{equation}%
We consider the static case, in which $S\partial _{t}S^{\dag }=0$. 
 Under the adiabatic approximation, the off-diagonal terms in Eq.(\ref{equnew}) can be omitted.  Consequently, Eq.(\ref{equnew}) can be  split into two decoupled equations,
\begin{eqnarray}
i\partial _{t}\tilde\Phi _{\pm } &=&\frac{1}{2m}\left( -i\nabla \mp \frac{\tilde{%
\Delta}}{\Lambda }\mathbf{k }_{r}\right) ^{2} \tilde\Phi _{\pm }  \notag \\
&&+\left(V_{\text{trap}}+\frac{Q}{2} \pm \frac{\Lambda }{2}\right) \tilde\Phi
_{\pm }  \label{Ad88} \\
&&+\frac{\Omega ^{2}}{8m\Lambda ^{2}}\left[ \left( \frac{\nabla \tilde{\Delta%
}}{\Lambda }-\frac{\tilde{\Delta}\nabla \Omega }{\Lambda \Omega }\right)
^{2}+4\mathbf{k}_{r}^{2}\right] \tilde\Phi _{\pm }.  \notag
\end{eqnarray}

 If the atom is in the lower $\left( -\right) $  branch,  then $\tilde\Phi _{+}=0$,  so we denote $\tilde \Phi _{-}$ as $%
\psi $ , and use (\ref{Ad7}) and (\ref{Ad77}),  giving
\begin{eqnarray}
\rho _{\uparrow } &=&\left\vert S_{21}\right\vert ^{2}\rho =\frac{1}{2}%
\left( 1-\tilde{\Delta}/\Lambda \right) \rho, \\
\rho _{\downarrow } &=&\left\vert S_{22}\right\vert ^{2}\rho =\frac{1}{2}%
\left( 1+\tilde{\Delta}/\Lambda \right) \rho,
\end{eqnarray}
where $\rho=|\psi|^2$, $\rho_\uparrow=|\Phi_\uparrow|^2$ and $%
\rho_\downarrow=|\Phi_\downarrow|^2$. Especially for the short-ranged interaction case, $%
V_{\sigma \sigma ^{\prime }}(\mathbf{r}_{i}-\mathbf{r}_{j})=g_{\sigma \sigma
^{\prime }}\delta (\mathbf{r}_{i}-\mathbf{r}_{j}),$ it has
\begin{eqnarray*}
G_{1}\left( \mathbf{r}\right) &=&\frac{1}{2}\left( 1-\tilde{\Delta}/\Lambda
\right) \rho g_{\uparrow \uparrow }+\frac{1}{2}\left( 1+\tilde{\Delta}%
/\Lambda \right) \rho g_{\uparrow \downarrow }\newline,
\\
G_{2}\left( \mathbf{r}\right) &=&\frac{1}{2}\left( 1+\tilde{\Delta}/\Lambda
\right) \rho g_{\downarrow \downarrow }+\frac{1}{2}\left( 1-\tilde{\Delta}%
/\Lambda \right) \rho g_{\uparrow \downarrow }.
\end{eqnarray*}%
Substituting the two equations into (\ref{Ad4}), we arrive at
\begin{eqnarray}
G &=&\frac{\rho }{2}\left[ g_{a}-{g_{as} \tilde \Delta }/{\Lambda}\right], \label{Gs} \\ 
Q &=&\frac{\rho }{2}\left[ 4g_{s}-g_{a}\tilde{\Delta}/\Lambda \right], \label{qq}
\end{eqnarray}%
where $g_{a}=g_{\uparrow \uparrow }-g_{\downarrow \downarrow },g_{s}=\left(
g_{\uparrow \uparrow }+g_{\downarrow \downarrow }+2g_{\uparrow \downarrow
}\right) /4$ and $g_{as}=\left( g_{\uparrow \uparrow }+g_{\downarrow
\downarrow }-2g_{\uparrow \downarrow }\right)$.  Eq.(\ref{Gs}) is a self-consistent equation of $G$, since $\tilde{\Delta}=\Delta+G$
and $\Lambda=\sqrt{\tilde\Delta^2+\Omega^2}$ are also functions of $G$.

For zero detuning $\Delta =0$ and in weak interaction limit, i.e., $\rho g_{a,s,as}\ll \Omega ,$ the only solution of Eq.(\ref{Gs}) and (\ref{qq}) is $G=%
\frac{1}{2}g_{a}\rho $ and $Q=2g_{s}\rho$ in the leading order, which also implies $\Lambda =\sqrt{\left( \Delta +G\right) ^{2}+\Omega ^{2}}\approx \Omega $.  Using the fact $\left( \frac{\nabla \tilde{\Delta}}{\Lambda }\right)
^{2}\ll 4\mathbf{k}_{r}^{2}$ \cite%
{juzeliunas20131}, and considering the case $\Omega $ being a constant (i.e., $%
\nabla \Omega =0$), Eq.(\ref{Ad88}) becomes
\begin{eqnarray*}
i\partial _{t}\psi &=&\frac{1}{2m}\left( -i\nabla +\frac{g_{a}\rho}{2\Omega }
\mathbf{k}_{r}\right) ^{2}\psi \\
&&+\left( V_{\text{trap}}+g_{s}\rho -\frac{\Omega }{2}\right) \psi +\frac{%
\mathbf{k}_{r}^{2}}{2m}\psi.
\end{eqnarray*}
Omitting those constant terms, we obtain the final form of GP equation in the weak interaction limit,
\begin{equation}
i\partial _{t}\psi =\frac{1}{2m}\left( -i\nabla +\frac{g_{a}\rho }{2\Omega }%
\mathbf{k}_{r}\right) ^{2}\psi +V_{\text{trap}}\psi +g_{s}\rho \psi .
\end{equation}


\section{Excitation energy in a uniform condensate}

For the uniform potential case, the GP equation is
\begin{equation}
i\partial _{t}\psi =\frac{1}{2m}(-i\nabla +\mathbf{a}\rho )^{2}\psi +b\rho
\psi,
\end{equation}%
where $\mathbf{a}=g_a \mathbf{k}_{r} /2\Omega$ and $b=g_s$. Taking $\psi =\sqrt{\rho }e^{i\eta }$, we obtain the Hydrodynamic equations,
\begin{eqnarray}
\partial _{t}\rho &=&-\nabla \cdot \left( \rho \mathbf{v}\right),  \label{hy1}
\\
\partial _{t}\eta &=&-\left( -\frac{1}{2m\sqrt{\rho }}\nabla ^{2}\sqrt{\rho }%
+\frac{m\mathbf{v}^{2}}{2}+b\rho \right) ,  \label{hy2}
\end{eqnarray}%
where $m\mathbf{v}=\nabla \eta +\mathbf{a}\rho .$

In the following, we consider the fluctuation of wavefunction around the static solution, i.e.,
\begin{eqnarray}
\rho &=&\rho _{0}+\delta \rho , \\
\eta &=&\eta _{0}+\delta \eta , \\
m\mathbf{v} &=&\nabla \eta _{0}+\mathbf{a}\rho _{0}+\nabla \delta \eta +%
\mathbf{a}\delta \rho.
\end{eqnarray}%
It is easy to find $m\mathbf{v}_{0}=\mathbf{\nabla }\eta _{0}+\mathbf{a}\rho
_{0}$ and $m\delta \mathbf{v=}\mathbf{\nabla }\delta \eta +\mathbf{a}\delta \rho .$
We start from the assumption that the static solution $\rho _{0}$ and $\mathbf{v}_{0}$ are constant
(vector) for uniform case. Thus, the linear perturbative expansion of Eq.(\ref{hy1}) is
\begin{equation}
\partial _{t}\delta \rho =-\nabla \delta \rho \cdot \mathbf{v}_{0}\mathbf{-}%
\frac{\rho _{0}}{m}\left( \mathbf{\nabla }^{2}\delta \eta +\mathbf{a}\cdot
\nabla \delta \rho \right) ,
\end{equation}%
which shows that the Laplace of $\delta\eta$ satisfies
\begin{equation}
\mathbf{\nabla }^{2}\delta \eta =-\nabla \delta \rho \cdot \frac{m\mathbf{v}%
_{0}}{\rho _{0}}\mathbf{-}\frac{m\partial _{t}\delta \rho }{\rho _{0}}-%
\mathbf{a}\cdot \nabla \delta \rho .  \label{hy3}
\end{equation}%
Similarly, the perturbative expansion of Eq.(\ref{hy2}) reads
\begin{equation}
\partial _{t}\delta \eta =-\left( -\frac{1}{4m\rho _{0}}\nabla ^{2}\delta
\rho +m\mathbf{v}_{0}\cdot \delta \mathbf{v}+b\delta \rho \right).
\label{hy4}
\end{equation}%

Perform time derivation of Eq.(\ref{hy3}) and use it to minus the Laplace of Eq.(\ref{hy4}), i.e., $\partial _{t}$(\ref{hy3})$-\mathbf{\nabla }^{2}$(\ref{hy4}), then we obtain
\begin{equation}
\begin{split}
\frac{\left( \nabla ^{2}\right) ^{2}}{4m\rho _{0}}& \delta \rho -\mathbf{v}%
_{0}\cdot \left( \nabla \nabla ^{2}\delta \eta +\mathbf{a}\nabla ^{2}\delta
\rho \right) -b\nabla ^{2}\delta \rho \\
& =-\partial _{t}\nabla \delta \rho \cdot \frac{m\mathbf{v}_{0}}{\rho _{0}}%
\mathbf{-}\frac{m\partial _{t}^{2}\delta \rho }{\rho _{0}}-\mathbf{a}\cdot
\partial _{t}\nabla \delta \rho,
\end{split}
\label{hy5}
\end{equation}%
where we have used the fact $m\delta \mathbf{v}=\mathbf{\nabla }\delta \eta +\mathbf{a}\delta
\rho $. Now substitute (\ref{hy3}) into (\ref{hy5}) to cancel $\nabla ^{2}\eta $,
then we get the differential equation of $\delta \rho $,
\begin{eqnarray}
&&{-}\partial _{t}^{2}\delta \rho -2\mathbf{v}_{0}\cdot \partial _{t}\nabla
\delta \rho -\frac{\mathbf{a}\rho _{0}}{m}\cdot \partial _{t}\nabla \delta
\rho  \notag \\
&=&\frac{\left( \nabla ^{2}\right) ^{2}}{4m^{2}}\delta \rho -\frac{b\rho _{0}%
}{m}\nabla ^{2}\delta \rho -\frac{\mathbf{v}_{0}\cdot \mathbf{a}\rho _{0}}{m}%
\nabla ^{2}\delta \rho  \notag \\
&&+\left( \mathbf{v}_{0}\cdot \nabla \right) ^{2}\delta \rho +\left( \mathbf{%
v}_{0}\cdot \nabla \right) \left( \frac{\mathbf{a}\rho _{0}}{m}\cdot \nabla
\right) \delta \rho.  \label{hy9}
\end{eqnarray}%
Substitute $\delta \rho \propto \exp (-i\omega t+i\mathbf{k\cdot r})$ into Eq.(\ref{hy9}), we find the
the excitation spectrum,
\begin{eqnarray}
\omega &=&\left( \mathbf{k}\cdot \mathbf{v}_{0}+\mathbf{k}\cdot \frac{%
\mathbf{a}\rho _{0}}{2m}\right)  \label{hy6} \\
&&+\sqrt{\frac{\mathbf{k}^{4}}{4m^{2}}+\frac{\mathbf{k}^{2}}{m}\left( b\rho
_{0}+\mathbf{v}_{0}\cdot \mathbf{a}\rho _{0}\right) +\left( \mathbf{k}\cdot
\frac{\mathbf{a}\rho _{0}}{2m}\right) ^{2}}.  \notag
\end{eqnarray}
Define $\mathbf{v}_{a}=\frac{\mathbf{a}\rho _{0}}{2m}$ and $c^{2}=\frac{%
b\rho _{0}}{m},$ and re-denote ${\bf{v}}_s\equiv\bf{v}_0$, then we obtain the formula
\begin{eqnarray}
\omega &=&\mathbf{k}\cdot \left( \mathbf{v}_{s}+\mathbf{v}_{a}\right) \notag\\
&&+\sqrt{\left[ c^{2}+2\mathbf{v}_{s}\cdot \mathbf{v}_{a}+\left( \hat{k}%
\cdot \mathbf{v}_{a}\right) \right] ^{2}\mathbf{k}^{2}+\frac{\mathbf{k}^{4}}{%
4m^{2}}},
\end{eqnarray}
which is used in the letter.
\section{Ground state wavefunction of a trap condensate}

Similar to the uniform potential case, the static Hydrodynamic equation in a harmonic trap is
\begin{eqnarray}
0 &=&-\nabla \cdot \left( \rho \mathbf{v}\right), \label{s31} \\
\mu &=& -\frac{1}{2m\sqrt{\rho }}\nabla ^{2}\sqrt{\rho }+\frac{m%
\mathbf{v}^{2}}{2}+b\rho +\frac{1}{2}m\omega ^{2}\mathbf{r}^{2} ,\label{s32}
\end{eqnarray}%
where $m\mathbf{v}=\nabla \eta +\mathbf{a}\rho .$ Use the Thomas-Fermi
approximation, and denote $\mu =\frac{1}{2}m\omega ^{2}R^{2}$, we obtain the density $\rho =\frac{1}{2}m\omega ^{2}\left(
R^{2}-\mathbf{r}^{2}\right)/b$ . The Thomas-Fermi radius  $R$ can be determined by the total number of particle.

Using Eq.(\ref{s31}),It is easy to get the equation of phase
\begin{equation}
-\rho \nabla ^{2}\eta -\nabla \rho \cdot \nabla \eta =\mathbf{a\cdot }\nabla
\rho ^{2}.
\end{equation}
Suppose $\mathbf{a}=a\hat{x}$ and substitute the density, then we have
\begin{equation}
-\frac{1}{2}\left( R^{2}-r^{2}\right) \nabla ^{2}\eta +r\partial _{r}\eta =f(\bfr), \label{s34}
\end{equation}
where $f(\bfr)=-a m \omega ^{2}x\left( R^{2}-r^{2}\right)/b$. To solve this differential equation, we need to solve the corresponding Homogeneous equation as following firstly,
\begin{equation}
-\frac{1}{2}\left( R^{2}-r^{2}\right) \nabla ^{2}\eta +r\partial _{r}\eta =\epsilon \eta. \label{s35}
\end{equation}

\begin{figure}
\includegraphics[width=5 cm]{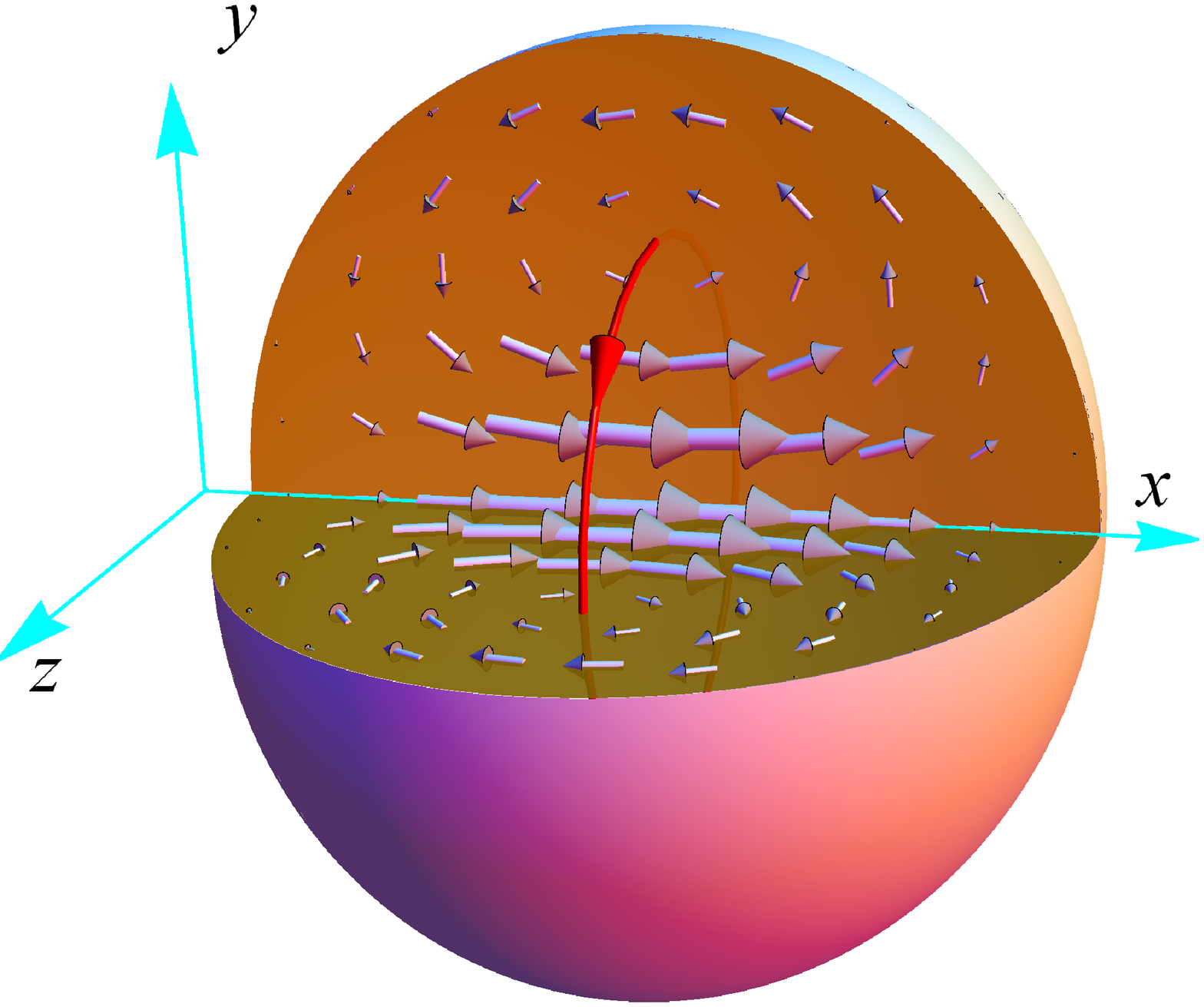}\newline
\caption{ The current (arrows) distribution in a 3D harmonic trap. The laser beams are along x direction, i.e., $\bfk_r=k_r \hat{x}$. The recoiled momentum is $k_r =2\pi\times 1 \mathrm{\mu m}^{-1}$ ($E_r \simeq 2 \pi \times 2.3 \mathrm{kHz}$) and the Raman coupling is $\Omega= 2 \pi \times 10 \mathrm{kHz}$.  We also choose the scattering length $a_a= 3.0 \mathrm{nm}$ and $a_s=6.0\mathrm{nm}$.  For total particle $N=1 \times 10^{6} $ in the trap $(\omega, \omega, \omega)=2 \pi \times (60, ~60, ~60)~ \mathrm{Hz}$, the Thomas-Fermi Radius  $R= 12.8 \mathrm{\mu m}$ and the corresponding chemical potential is $\mu \simeq 2 \pi \hbar \times 2.5 \mathrm{kHz}$. The radius of the coreless ring is $7.4 \mathrm{\mu m}$, and the direction on the ring is used to mark the direction of magnetic field $\mathbf{B}=\nabla \times \mathbf{a} \rho$.
}
\label{3dcurrent}
\end{figure}

For 2D case, use the polar
coordinate $\left( r,\theta \right) $ and decompose the phase function into $\eta
\left( r,\theta \right) =D(r)e^{-i l \theta }$, then Eq.(\ref{s35}) becomes
\begin{eqnarray}
\epsilon D\left( r\right)  &=&-\frac{\left( R^{2}-r^{2}\right)}{2} \left[
D^{\prime \prime }(r)+\frac{1}{r}D^{\prime }\left( r\right) -\frac{l ^{2}}{%
r^{2}}D\left( r\right) \right] \notag \\
&&+rD^{\prime }\left( r\right).
\end{eqnarray}
Subsequently, define a new radical function $G(r)=D\left( r\right) /r^{\left\vert l\right\vert }$, and substitute it into the above equation, then we have
\begin{eqnarray}
\epsilon G\left( r\right)  &=&-\frac{\left( R^{2}-r^{2}\right)}{2} \left[
G^{\prime \prime }(r)+\frac{2\left\vert l\right\vert +1}{r}G^{\prime }\left(
r\right) \right]  \notag\\
&&+\left\vert l\right\vert G\left( r\right) +rG^{\prime }\left( r\right). \label{Gfun}
\end{eqnarray}
Now, introduce the new variable $u=r^{2}/R^{2}$,  we can transform Eq.(\ref{Gfun}) into a familiar form
\begin{eqnarray}\label{nst}
0 &=&u\left( 1-u\right) G^{\prime \prime }(u)+\left[ \left( \left\vert
l\right\vert +1\right) -\left( \left\vert l\right\vert +2\right) u\right]
G^{\prime }\left( u\right)  \notag\\
&&+\frac{\epsilon -\left\vert l\right\vert }{2}G\left( u\right), \label{s37}
\end{eqnarray}
which is the standard form for the Hypergeometric function $_{2}F_{1}\left( \alpha ,\beta ,\gamma ;u\right)$,
\begin{equation}\label{st}
0=u\left( 1-u\right) F^{\prime \prime }\left( u\right) +\left[ \gamma
-\left( \alpha +\beta +1\right) u\right] G^{\prime }\left( u\right) -\alpha
\beta G\left( u\right).
\end{equation}

For the function to be well behaved, it requires $\alpha =-n,$ where $n$ is non-negative integer. Comparing (\ref{nst}) with (\ref{st}),
we have $\beta =\left\vert l\right\vert +n+1,\gamma =\left\vert l\right\vert
+1,\epsilon =\left\vert l\right\vert +2n+2n\left\vert l\right\vert +2n^{2}.$ Conclusively, the eigenfunction of (\ref{s35}) is   $\eta_{nl}= \tilde{C}_{nl} r^{|l|}{}_{2}F_{1}\left( -n,\left\vert l\right\vert
+n+1,\left\vert l\right\vert +1;\frac{r^{2}}{R^{2}}\right) e^{-il\theta}$ with eigenvalue
$\epsilon_{nl}= \left\vert l\right\vert
+2n+2n\left\vert l\right\vert +2n^{2} .$ Here, $\tilde{C}_{nl}$ is the normalization coefficient.

Now we can solve Eq.(\ref{s34}) by Green's function method,
\begin{equation}  \label{phase3dsolution}
\eta(\mathbf{r}) = \sum_{n,l} \int d\mathbf{r}^{\prime }\eta_{nl}(\mathbf{r%
})\frac{1}{\epsilon_{nl}} \eta_{nl}^*(\mathbf{r^{\prime }})f(\mathbf{r}%
^{\prime })   \\
\end{equation}
and get the phase $\eta(\bfr)$.

For 3D case, there is a detailed calculation of solving differential Eq.(\ref{s35}) in Ref.\cite{Smith2008} and the process is similar to 2D case, so we will not repeat the detail here but show the results.  The eigenfunction of Eq.(\ref{s35}) in 3D is $\eta_{nlm}= {C}_{nl} r^l{}_{2}F_{1}\left( -n,l+n+3/2,l+3/2;\frac{r^{2}}{R^{2}%
}\right)Y_{lm} $ with eigenvalue $\epsilon_{nlm}= l+3n+2nl+2n^{2} .$ Here ${C}_{nl}$ is the normalization coefficient. Using the Green function method, we can get the phase $\eta(\bfr)$ and the corresponding velocity shown in the letter.
In Fig.\ref{3dcurrent}, we show the current
distribution in a 3D trap, where the persistent current forms a coreless
vortex ring with radius $R/\sqrt{3}$, which circulate the $x$ axis (parallel to the gauge field
direction).